# Unified Field Multiplier for ECC: Inherent Resistance against Horizontal SCA Attacks

Ievgen Kabin, Zoya Dyka, Dan Kreiser and Peter Langendoerfer

*IHP*
*Im Technologiepark 25*
*Frankfurt (Oder), Germany*

*Abstract*—In this paper we introduce a unified field multiplier for the EC *kP* operation in two different types of Galois fields. The most important contributions of this paper are that the multiplier is based on the 4-segment Karatsuba multiplication method and that it is inherent resistant against selected horizontal attacks.

## I. Introduction

Elliptic Curve Cryptography (ECC) can guarantee not only confidentiality of communication but can also be used for authentication of persons/devices. Nowadays ECC is worldwide implemented for use in the Internet of Things, WSNs, automation industry, protection of critical infrastructures, etc. The private key has to be kept secret; the applied cryptographically strong cipher algorithm, its input and output values may be publicly known. So, the goal of attackers is to reveal the private key. Cryptographic algorithms are implemented either in software or in hardware, i.e. they run on a device or are part of it. The current through the device, its electromagnetic radiation and other physical parameters are caused by each execution of cryptographic operations. If an attacker has physical access to the device, he can exploit those side channel effects to reveal its private key. Misusing the key can cause dangerous situations in traffic, e.g. car crashes, etc. In order to avoid successful malicious attacks the cryptographic implementations need to be protected against side channel analysis (SCA) attacks.

In 2010 Clavier et al. classified SCA attacks into vertical and horizontal [1], using the number of measured traces as classification criterion. Horizontal attacks are single-trace attacks; vertical attacks are "more than one trace" attacks: the attacker measures and analyses more than one trace of a crypto-execution, with different inputs. The number of traces needed can vary between just two up to several millions.

Well-known countermeasures such as elliptic curve (EC) point blinding, randomization of projective coordinates of EC points or the key randomization proposed in [2] are effective against vertical attacks but not against horizontal attacks.

Designing a unified accelerator for EC point multiplication is a not trivial task, especially if the design has to be resistant against horizontal SCA attacks. An EC can be constructed over finite fields that are used in cryptographic applications due to their suitability for implementation in digital systems. Especially the prime field *GF(p)* and binary extension field $GF(2^n)$ are favorable since various standardization bodies such as NIST and ANSI specifically recommend several elliptic curves over these finite fields. A hardware implementation for performing multiplications in both types of fields using the same data path (i.e. a unified architecture) were discussed in literature. In most of the cases they are designed and analyzed from the time and area efficiency and complexity points of view. However, their resistance against SCA attacks are still not investigated so far. To the best of our knowledge this is the first attempt of a unified field multiplier implementation that is resistant against known horizontal SCA attacks, particularly against attacks introduced in [1], [3]-[5].

## II. Our Basic Design

In this section we explain the implementation details of the design that we use as a basis for implementing a unified *kP* accelerator for the following 4 NIST ECs: *P-224*, *P-256*, *B-233* and *B-283*. Our basic design is a *kP* accelerator for NIST EC *B-233* only [6]. We implemented a modified Montgomery *kP* algorithm using Lopez-Dahab projective coordinates [7] as proposed in [8] to prevent revealing of the second most significant bit of the scalar *k*. The implemented algorithm is regular. Furthermore, we applied the *side-channel atomicity* principle i.e. we ensured that the processing of each key bit is implemented using the same operation sequence, including also the write to register operations in the internal blocks of our design. In addition to both SPA countermeasure approaches – the *regularity* and the *atomicity* principle – we increased the inherent resistance of our design by performing additions, squarings and write to register operations always in parallel to the field multiplication.

The multiplication is the most complex field operation. In order to tackle this complexity issue many new multiplication formulae or their combinations have been proposed in the past. Many multiplication methods (MM) apply segmentation of both multiplicands into the same number of parts. We implemented our field multiplier using the 4-segment Karatsuba MM according to a fixed calculation plan as described in [9]. Our design uses only one field multiplier that calculates the field product of 233 bit long operands. One field multiplication takes only 9 clock cycles. Note that the sequence of partial product calculations given in [9] is not the only one leading to correct results. All partial products are accumulated in a register of the multiplier. The field product will be accumulated iteratively, clockwise, using the calculated partial products. The reduction is also performed in each clock cycle.

### III. Performed horizontal attacks

In this section we describe the horizontal attacks that we performed against our *kP* design and its field multiplier. We synthesized the *kP* design using the gate library for the IHP 250 nm technology. We simulated the power consumption of the *kP* design while the *kP* operation was executed with a randomly chosen EC point *P* and a randomly generated 232 bit long scalar *k*. The design was simulated using the Synopsis PrimeTime suit. The power traces were simulated for a clock cycle of 30 ns. Because the simulated traces are noiseless and no data are lost in simulations, we represented each clock cycle using only one power value – i.e. the average power value of the clock cycle.

#### A. Horizontal attack using the difference of the mean

We performed a horizontal DPA attack using the *difference of the mean* as described in [5]. We analysed power profiles of iterations of the main loop of the algorithm. The corresponding part of a power traces is denoted as a slot in the rest of this paper. So in our case we analysed 230 slots, each corresponding to the processing of a single key bit. The processing of a key bit in the main loop takes only 54 clock cycles, i.e. each compressed slot consists of 54 values (samples). We calculated the average slot and we compared it pointwise with each of the 230 power profiles. In the following *j* denotes the slot within a power trace, i.e. $0 \leq j \leq 229$, while *i* denotes the clock cycle within a slot i.e. $1 \leq i \leq 54$. For each point of the average slot we obtained a 230 bit long key candidate using the following assumption: the $j^{th}$ bit of the key candidate is 1 if in the slot with number *j* the power value with number *i* is smaller than or equal to the power value with number *i* in the average slot. Else the $j^{th}$ bit of the $i^{th}$ key candidate is 0. For each key candidate we calculated its relative correctness *δ* in per cent. We inverted each key candidate with a correctness *δ*<50%, because it means that our assumption was not true, i.e. the opposite assumption is correct. Thus, we obtained all key candidates with a correctness in the range between 50% and 100%. **Fig. 1** shows the correctness of the key candidates. 20 of 54 key candidates have a correctness between 70% and 100% if the power consumption of the whole design is analysed (see red solid line). In contrast to the attack results that we obtained analysing power traces of whole *kP* design, the same attack using the power trace of the single block MULT, i.e. the field multiplier, only, was not successful, see blue dotted line.

The strong information leakage source in our *kP* design is its bus. We learned this by analyzing the power traces of single blocks of the *kP* design. Although we implemented the *kP* design strongly balanced and the field multiplier is always active, the addressing of registers and other blocks depending on the processed scalar *k* can cause its successful extraction.

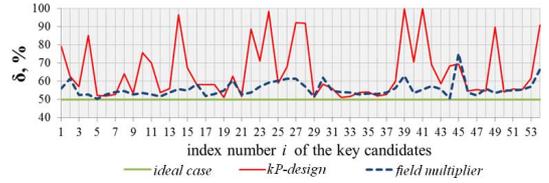

**Fig. 1.** Result of the horizontal attack using *the difference of the mean* for statistical analysis of simulated Power Traces of the *kP* design (red solid line) and its field multiplier (blue dotted line).

#### B. Bauer attack

In [4] a Horizontal Collision Correlation Attack on ECs was published. The main assumption is that an attacker can distinguish two multiplications with at least one common multiplicand from two multiplications with different multiplicands. This knowledge can be used for revealing the key because a point doubling can be distinguished from a point addition even in double-and-add *kP* algorithms obeying the atomicity principle [10]. The field multiplier analysed in the attack in [4] was implemented using the classical MM. Experimental results in [4] show that Pearson's coefficients calculated for traces of two multiplications with a common operand differ significantly from coefficients calculated for two multiplications with different operands. Using this type of attack against the Montgomery *kP* algorithm will not reveal the key but can help to separate the trace into slots. The fact that a multiplication with EC parameter *b* (or a multiplication with the *x* coordinate of input point *P*) is executed in each slot can be exploited to segment the power trace into slots. We performed a horizontal collision correlation attack as follows:

*1)* We calculated an average power profile of the multiplication with the parameter *b* using profiles of all multiplications with this operand within the *kP* operation.

*2)* We calculated Pearson's coefficients for the averaged power profile and each multiplication within the processing of the scalar *k* in the main loop of the implemented algorithm.

*3)* We represented the calculated coefficients graphically, see **Fig. 2**-*a*. The simulated trace contains 230·6=1380 field multiplications performed in the main loop of the implemented algorithm. In **Fig. 2**-*a* the Pearson's coefficients for the averaged power profile and a multiplication with the operand *b* are marked red, all others are blue.

The result of the analysis is: if a trace of the whole *kP* design is analysed the calculated Pearson's coefficients allow to distinguish in most but not in all cases multiplications with the operand *b* (see red dots in **Fig. 2**-*a*) from other multiplications (see blue dots in **Fig. 2**-*a*). We obtained similar results using the average power profile of multiplications with the *x* coordinate of input point *P* as well as using the average power profile of a multiplication with always different operands. In our experiments they are the 1st and the 3rd multiplication in each slot, respectively. We assumed that not the field multiplier but other operations performed in parallel to the multiplication cause this effect. To proof this idea we analysed the power trace of the single block MULT of the $kP$ design in the same way (steps 1-3) as the power trace of the complete design. If a trace of the field multiplier, only, is analysed (see **Fig. 2**-b) multiplications with the operand $b$ (marked as red dots) cannot be distinguished from other multiplications (marked blue).

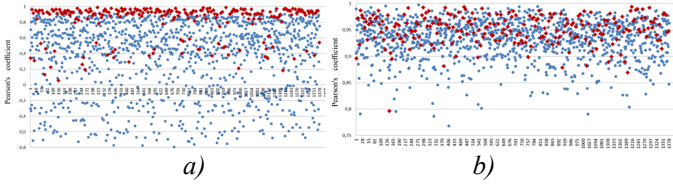

**Fig. 2.** Result of the horizontal collision correlation analysis: *a)* a trace of the whole *kP* design is analysed; *b)* a trace of the field multiplier, only, is analysed

## C. Inherent resistance of investigated field Multiplier

Please note, that the traces analysed here are noiseless, i.e. the signal/noise ratio is equal to infinity. Even in this case (it is the ideal case in [4]) the horizontal attacks described above were not successful i.e. the used scalar $k$ was not revealed using *the difference of the mean* and multiplications with a common operand are not distinguishable from other multiplications (see subsection III-B). We implemented the field multiplier without any countermeasures against SCA, i.e. the sequence of the multiplication was not randomized as for example in [11], and multiplications are not masked. The investigated multiplier is inherently resistant against the performed attacks. Due to this fact we propose to design a unified field multiplier, for ECs *P-224, P-256, B-233* and *B-283*, based on the 4-segment Karatsuba MM. To the best of our knowledge such a multiplier was not yet reported in literature. In comparison to a multiplier based on the classical MM, the proposed multiplier is faster (9 clock cycles in comparison to 16 that needed by a multiplier applying the classical MM using the same segmentation of operands), and is inherently resistant against horizontal attacks.

## IV. PROPOSED UNIFIED FIELD MULTIPLIER

We represent the 4-segment Karatsuba MM (1) for large binary integer numbers $A$ and $B$ as **Table 1**:

$$A \cdot B = (A_3 \cdot 2^{3m} + A_2 \cdot 2^{2m} + A_1 \cdot 2^{1m} + A_0) \cdot (B_3 \cdot 2^{3m} + B_2 \cdot 2^{2m} + B_1 \cdot 2^{1m} + B_0) =$$
$$= s_6 \cdot 2^{6m} + s_5 \cdot 2^{6m} + s_4 \cdot 2^{4m} + s_3 \cdot 2^{3m} + s_2 \cdot 2^{2m} + s_1 \cdot 2^{1m} + s_0 =$$
$$= \begin{pmatrix} p_1 \cdot 2^0 + p_7 \cdot 2^{1m} + (p_2 + p_5) \cdot 2^{2m} + \\ + (p_1 + p_2 + p_3 + p_4 + p_9) \cdot 2^{3m} + \\ + (p_3 + p_6) \cdot 2^{4m} + p_8 \cdot 2^{5m} + p_4 \cdot 2^{6m} \end{pmatrix} - \begin{pmatrix} (p_1 + p_2) \cdot 2^{1m} + (p_1 + p_3) \cdot 2^{2m} + \\ (p_5 + p_7 + p_6 + p_8) \cdot 2^{3m} + \\ + (p_2 + p_4) \cdot 2^{4m} + (p_3 + p_4) \cdot 2^{5m} \end{pmatrix} = (1)$$
$$= C_7 C_6 C_5 C_4 C_3 C_2 C_1 C_0,$$

with

$p_1 = A_0 B_0,\ p_2 = A_1 B_1,\ p_3 = A_2 B_2,\ p_4 = A_3 B_3,$ } each $2m$ bits long
$p_5 = (A_0 + A_2)(B_0 + B_2),\ p_6 = (A_1 + A_3)(B_1 + B_3),$
$p_7 = (A_0 + A_1)(B_0 + B_1),\ p_8 = (A_2 + A_3)(B_2 + B_3),$ } each $2(m+1)$ bits long
$p_9 = (A_0 + A_1 + A_2 + A_3)(B_0 + B_1 + B_2 + B_3)$ } $2(m+2)$ bits long

each segment of the product $C_i$ is $m$ bits long.

**Table 1.** Representation of formula (1).

| | $C_7$ | $C_6$ | $C_5$ | $C_4$ | $C_3$ | $C_2$ | $C_1$ | $C_0$ | | |
|---|---|---|---|---|---|---|---|---|---|---|
| $p_1$ | | | | + | | + | | | $p_{1(+)}$ | $clk^5$ |
| | | | | | - | | | | $p_{1(-)}$ | |
| $p_2$ | | | | | + | | | | $p_{2(+)}$ | $clk^6$ |
| | | | | - | | - | | | $p_{2(-)}$ | |
| $p_3$ | | | | + | | | | | $p_{3(+)}$ | $clk^9$ |
| | | | - | | - | | | | $p_{3(-)}$ | |
| $p_4$ | | + | | + | | | | | $p_{4(+)}$ | $clk^1$ |
| | | | | - | | | | | $p_{4(-)}$ | |
| $p_5$ | | | | + | | | | | $p_{5(+)}$ | $clk^8$ |
| | | | | | - | | | | $p_{5(-)}$ | |
| $p_6$ | | | + | | | | | | $p_{6(+)}$ | $clk^3$ |
| | | | | - | | | | | $p_{6(-)}$ | |
| $p_7$ | | | | | | + | | | $p_{7(+)}$ | $clk^7$ |
| | | | | | - | | | | $p_{7(-)}$ | |
| $p_8$ | | | + | | | | | | $p_{8(+)}$ | $clk^2$ |
| | | | | - | | | | | $p_{8(-)}$ | |
| $p_9$ | | | | | + | | | | $p_{9(+)}$ | $clk^4$ |
| | $C_7$ | $C_6$ | $C_5$ | $C_4$ | $C_3$ | $C_2$ | $C_1$ | $C_0$ | | |

The representation of formula (1) as a table consists of all partial products $p_j$ (see the most left column of **Table 1**), and shows how the partial products have to be accumulated, i.e. to which $m$ bit long segment $C_i$ and by which operation – addition ('+' in green cells) or subtraction ('–' in yellow cells). Each summand consisting of the partial products $p_j$ to be added to the result is denoted by $p_{j(+)}$; each summand consisting of the partial products $p_j$ to be subtracted from the result is denoted by $p_{j(-)}$. The fact that the partial products $p_5, p_6, p_7$ are $2m+2$ bits long is represented by grey cells in the most left column in **Table 1**. The $2m+4$ bits long partial product $p_9$ is marked in black. The last column of **Table 1** gives our sequence of calculation of the partial products. The cells indicate the clock cycle $clk^j$, $1 \leq j \leq 9$, in which the operation shown in the corresponding row is executed. Please note that all additions and subtractions in formula (1) and **Table 1** are bitwise XOR operations for product calculation in $GF(2^n)$. In order to adapt the 4-segment Karatsuba MM for calculating products for $GF(p)$ as well as for $GF(2^n)$ we applied the Carry Bit separation (CBS) technique for obtaining the operands for the Partial Multiplier, for calculating each partial product $p_j$, as well as for their accumulation corresponding to **Table 1**. The structure of our unified field multiplier is shown in **Fig. 3**.

We applied the CBS technique for all additions and partial multiplications. The CBS technique takes the fact in account that the addition in $GF(2^n)$ is a bitwise XOR operation. All carry values can be calculated separately. The sum in $GF(p)$ is the sum of the sum in $GF(2^n)$ and the carry values. The calculation of a partial product applying the CBS technique can be described as: $p = p^{XOR} + sel \cdot p^{carry}$. Here $sel$ is '1' if product in $GF(p)$ has to be calculated and '0' for the product calculation in $GF(2^n)$.

We described the functionality of our unified polynomial multiplier in VHDL and synthesized it for the IHP 250 nm technology. Parameters of the unified field multiplier as well as performance data such as the area are given in **Table 2**. For comparison **Table 2** contains additionally data about the field multiplier for EC *B-233* that is a block in our basic design.

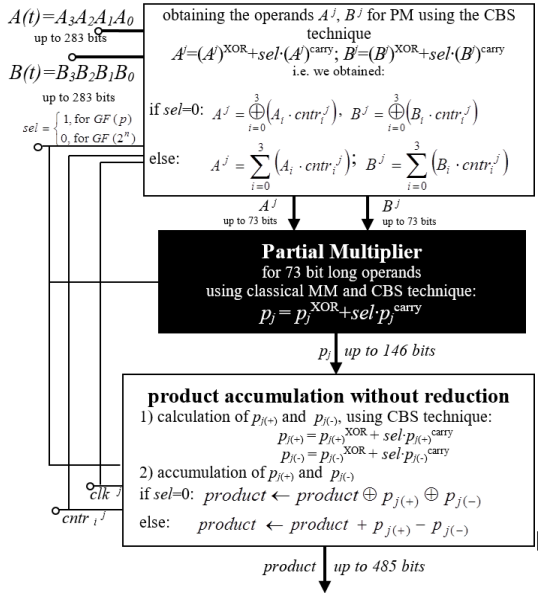

**Fig. 3.** Structure of the unified field multiplier (without reduction) implemented using the 4-segment Karatsuba MM applying the CBS technique.

**Table 2.** Parameters of synthesized multipliers.

|  | *B-233*: field multiplier | our unified multiplier |
|---|---|---|
| MM | 4-segment Karatsuba | 4-segment Karatsuba |
| area | 0.5 mm$^2$ | 3.2 mm$^2$ |
| clock cycle | 30 ns | 100 ns |
| power | 22.4 mW | 61.3 mW for *B-233* (min) <br> 84.2 mW for *P-256* (max) |
| Partial Multiplier | - MM : area-optimized combination of 3 MMs; <br> - inputs: 59 bit long; <br> - area: 0.15 mm$^2$; <br> - power: 6.13 mW. | - MM: classical MM applying CBS technique; <br> - inputs: 73 bit long; <br> - area: 1.785 mm$^2$; <br> - power: 43.1 mW. |

To investigate the resistance of our unified multiplier against Horizontal Collision Correlation Attack (HCCA) introduced by A. Bauer et. al. in [4] we performed the following experiments:

- We simulated the power traces of the product calculation for 4 multiplications with one common and two completely different operands: $mult_1 = a \cdot b$; $mult_2 = c \cdot d$; $mult_3 = a \cdot e$; $mult_4 = f \cdot g$.

- We calculated Pearson coefficients $k_i$ ($1 \leq i \leq 4$) using the power shape of following multiplications: $k_1$ using $mult_1$ and $mult_3$, here operand *a* is common in 2 multiplications; $k_2$ using $mult_2$ and $mult_4$, $k_3$ using $mult_1$ and $mult_2$ and $k_4$ using $mult_1$ and $mult_4$. Coefficients $k_2$, $k_3$ and $k_4$ correspond to multiplications with different operands.

If the coefficient $k_1$ differs significantly from $k_3$ and/or $k_4$, the multiplications with a common operand $mult_1$ and $mult_3$ are distinguishable from multiplications with different operands such as $mult_1$ and $mult_2$ and/or $mult_1$ and $mult_4$. Please note that this distinguishability has to be observed for each experiment (see steps 1) and 2)) for a successful HCCA attack.

We performed the experiment described above for multiplicands of length 224 bits (for EC *P-224*), 233 bits (for EC *B-233*), 256 bits (for EC *P-256*) and for 283 bits (for EC *B-283*) 20 times each. The operands were randomly generated 283 bit long numbers. For operands of smaller length *n* we used the least significant *n* bits of the 283 bit long numbers. **Fig. 4** shows coefficients $k_1$, $k_2$, $k_3$ and $k_4$ calculated for all experiments. The coefficients $k_1$ are represented by red dots, $k_2$ by blue dots, $k_3$ by blue triangles and $k_4$ by blue crosses. It can be seen that the set of coefficients represented with red points is indistinguishable from other coefficients, i.e. our unified multiplier is resistant against HCCA attacks.

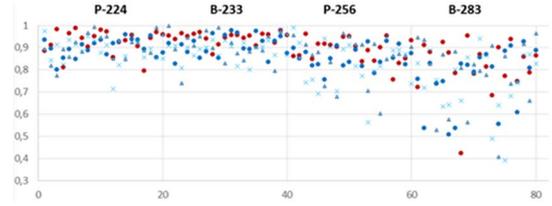

**Fig. 4.** Calculated Pearson coefficients (on *y*-axis): 4 coefficients per experiment (a value on *x*-axis); 20 experiments for each of the 4 investigated ECs.

ACKNOWLEDGMENT

The work presented here was supported by the Federal Ministry of Education and Research (BMBF) of Germany under grant number 03ZZ052717.